\documentstyle[twoside,fleqn,npb,epsfig]{article}

\title{Two Loop Electroweak Bosonic Corrections to the Muon Decay Lifetime}

\author{M. Awramik\address{Department of Field Theory and Particle Physics,
        Insitute of Physics, University of Silesia, \\
        Uniwersytecka 4, PL-40007 Katowice, Poland}
        and
        M. Czakon$^{\textrm{{\scriptsize{a}}},}$\address{Institut f\"ur Theoretische Physik, 
        Universit\"at Karlsruhe, \\
        D-76128 Karlsruhe, Germany}
\thanks{Presented at 6th International Symposium on Radiative Corrections: Application of Quantum Field Theory
Phenomenology (RADCOR 2002) and 6th Zeuthen Workshop on Elementary Particle Theory (Loops and Legs in
Quantum Field Theory), Kloster Banz, Germany, 8-13 Sep 2002.}
}

\begin{document}

\begin{abstract}
A review of the calculation of the two loop bosonic corrections to
$\Delta r$ is presented. Factorization and matching onto the Fermi model
are discussed. An approximate formula, describing the
quantity over the interesting range of Higgs boson mass values from
100~GeV to 1~TeV is given.
\end{abstract}

\maketitle

The muon decay lifetime ($\tau_\mu$) has been used for long as an
input parameter for high precision predictions of the Standard Model
(SM). It allows for an indirect determination of the mass of the $W$
boson ($M_W$), which suffers currently from a large experimental error
of 33~MeV \cite{Parkes:2002vv}, one order of magnitude worse than that
of the $Z$ boson mass ($M_Z$). A reduction of this error by LHC to
15~MeV \cite{lhctdr} and by a future linear collider to 6~MeV
\cite{tesla-tdr} would provide a stringent test of the SM by
confronting the theoretical prediction with the experimental value.

The extraction of $M_W$ with an accuracy matching that of next
experiments, {\it i.e.} at the level of a few MeV necessitates
radiative corrections beyond one loop order. Large two loop
contributions from fermionic loops have been calculated in
\cite{Freitas:2000gg}.  The current prediction is affected by two
types of uncertainties. First, apart from the still unknown Higgs
boson mass, two input parameters introduce large errors. The current
knowledge of the top quark mass results in an error of about 30~MeV
\cite{Freitas:2002ja}, which should be reduced by LHC to 10~MeV and by
a linear collider even down to 1.2~MeV. The inaccuracy of the
knowledge of the running of the fine structure constant up to the
$M_Z$ scale, $\Delta \alpha(M_Z)$, introduces a further $6.5$~MeV
error. Second, several higher order corrections are unknown. In fact
the last correction of order ${\cal O}(\alpha^2)$, generated by purely
electroweak bosonic loops, has been calculated only recently in
\cite{Awramik:2002wn} and confirmed by an independent group in
\cite{Onishchenko:2002ve}. The details of both calculations are
presented in \cite{Awramik:2002vu}.

The two loop bosonic corrections to the muon lifetime have been
previously estimated by two different methods. First, the leading term
in the large Higgs boson mass expansion has been obtained in
\cite{Halzen:1991ik} and \cite{jeg}. Although the results were
different, the size of the contributions was negligible.  Apart from
the disagreement, the large mass expansion is not justified for small
Higgs boson masses, and the experimental data seems to favour a low
mass range. Second, a resummation formula has been used
\cite{Freitas:2002ja}, which with respect to bosonic corrections
consists of assuming a geometric progression of the contributions with
the order of perturbation theory. This means that at two loop order,
the bosonic corrections should be equal to the square of the one loop
result. This, however, is contradicted by the fact, that the leading behaviour
with the Higgs boson mass would then be of logarithmic type ($\sim
\log^2(M_H^2)$), whereas it is known that the result should behave as
the square of this mass \cite{Veltman:1976rt}.

Since both estimates turn out to be unreliable the full calculation was
necessary. Here we shall describe the crucial ingredients, starting
from the proper definition of $\Delta r$, passing through the various
stages of the renormalization procedure, and ending with the methods
used in numerical evaluation.

The muon decay is described by an effective field theory, the Fermi
model, which consists of $QED$, $QCD$ of five light quarks and the four
fermion point interaction
\begin{equation}
  \label{LF}
  {\cal L}_{F} = \frac{G_F}{\sqrt{2}}\bar{e} \gamma^\alpha (1-\gamma_5) \mu
  \otimes \bar{\nu_\mu} \gamma_\alpha (1-\gamma_5) \nu_e.
\end{equation}
The constant $G_F$ should be predicted by the Standard Model. The
difference between the tree level prediction and the full result is
factorized in the quantity $\Delta r$
\begin{equation}
  \label{CF}
  \frac{G_F}{\sqrt{2}} = \frac{\pi
  \alpha}{2 M_W^2 s_W^2}(1+\Delta r).
\end{equation}
In this sense $\Delta r$ can be considered as the matching or Wilson
coefficient in the Fermi model. The matching equation assumes the form
\begin{equation}
S^{\rm SM} = S^{\rm Fermi}
+{\cal O}\left(\frac{p^4}{M_W^4},\frac{m^4}{M_W^4}\right),
\end{equation}
where $p$ and $m$ denote the external momenta and the light masses
respectively, and $S$ is the renormalized scattering matrix in both
models. The easiest way to obtain $\Delta r$ is to simply put $p$ and
$m$ equal to zero on both sides of the equation. This procedure will
of course generate spurious infrared divergences, which however will
cancel from the equation as has been shown in \cite{Gorishnii:dd}. The
main advantage comes from the fact, that on the right side, all loop
diagrams will be scaleless (as they contain only light particles), and
will therefore vanish, leaving only the tree level diagram trivially
proportional to $G_F$. Additionally, the SM diagrams on the left hand
side will reduce to vacuum bubbles.

\begin{figure}
\psfig{figure=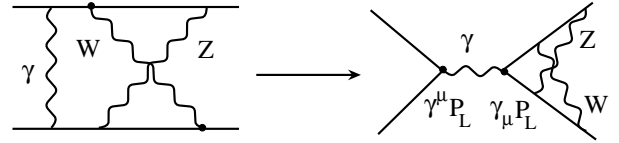,width=8cm}
\caption{\label{Fierz} Example of a Fierz rearrangement about the
  photon line.}
\end{figure}

A major point in the above procedure is connected with the treatement
of spinor chains. Dimensional regularization for spurious infrared
divergencies forbids a direct use of the Chisholm identity
\begin{equation}
  \gamma_\mu \gamma_\nu \gamma_\rho = g_{\mu\nu} \gamma_\rho +
  g_{\nu\rho} \gamma_\mu - g_{\mu\rho} \gamma_\nu -i
  \epsilon_{\mu\nu\rho\sigma} \gamma^\sigma \gamma_5
\end{equation}
for all diagrams, which are divergent after renormalization. A priori
we would have to introduce a whole basis of tensor products of
possible spinor chains. It turns out, however, that the projection onto
the Fermi operator can be
defined unambiguously through Fierz symmetry. The problematic diagrams
contain always a photon line connecting the two chains. A Fierz
rearrangement about this line or the $W$ boson line, as depicted in
Fig.~\ref{Fierz}, leads to an
object which has the required vertex structure
\begin{equation}
\label{basicOP}
\gamma^\mu P_L \otimes \gamma_\mu P_L.
\end{equation}
In practice, this Fierz rearrangement can be realised with the help of
a projector. In the present case, the transformation
\begin{equation}
\Gamma_1 \otimes \Gamma_2 \rightarrow -\frac{1}{2 D (D-2)} \mbox{Tr} (\Gamma_1
  \gamma_\mu P_R \Gamma_2 \gamma^\mu P_R)
\end{equation}
gives directly the coefficient of the operator Eq.~\ref{basicOP} in
$D$ dimensions.

The matching procedure requires proper renormalization of the
model. Although in the end we are interested in the on-shell
parametrization for the masses and the electric charge, an
intermediate renormalization can be chosen at will. In fact the
factorization theorem \cite{Gorishnii:dd} has been proven for mass
independent schemes and the on-shell scheme does not have this
property. However, as long as the parameters are defined at the heavy
scale the mass dependence does not contradict the theorem. For this
reason, we may safely renormalize the boson masses in the on-shell
scheme. The situation is somewhat more problematic with the electric
charge, which is usually defined by the Thompson scattering processes
for on shell electrons and photon. If we are interested in bosonic
corrections only, the definition can be kept, since an inspection of
the diagrams leads to the conclusion that no spurious infrared
divergencies can be introduced. For light fermionic contributions the
situation is different with the usual solution consiting of a shift of
the definition to the $M_Z$ scale \cite{Sirlin:1980nh}.

In this work, two renormalisation schemes have been used. First, the
complete calculation has been performed in the on-shell
scheme. Second, the model has been renormalized in the $\overline{\rm
  MS}$ scheme, and the result translated back to the on-shell scheme
by means of relations between the on-shell and $\overline{\rm MS}$
masses, which for the $W$ and $Z$ masses were required up to two loop
order. Obviously the translation should be applied to a scheme
independent quantity. In this case, this is the Fermi constant $G_F$,
and the correct relation is
\begin{equation}
\left.\frac{\pi\alpha}{2 M_W^2 s_W^2}(1+\Delta r) \right|_{\rm OS} = 
\left.\frac{\pi\alpha}{2 M_W^2 s_W^2}(1+\Delta r)
\right|_{\overline{\rm MS}}.
\end{equation}
It is not trivial that the formulae will coincide, since the on-shell
counterterms on the left hand side contain terms of order $\epsilon$
($\epsilon = 1/2(4-D)$), which are not used in the translation between
the schemes on the right hand side.

An interesting part of the definition of the model is the treatement
of tadpole diagrams. It is known that gauge invariance of mass
counter-terms requires inclusion of tadpoles \cite{appelquist,tadpole}
(at the two loop level this has been explicitely shown in
\cite{Jegerlehner:2001fb}). In this case, however, one cannot use
one-particle-irreducible (1PI) Green functions. In order to have gauge
invariant counter-terms and 1PI Green functions only, a special
procedure was designed. An additional renormalisation constant for the
bare vacuum expectation value $v_0$, denoted $Z_v$, has been
introduced and explicitely split from the bare masses
\begin{eqnarray}
  v_0 & \longrightarrow & v_0 Z_v^{1/2}, \\ 
  M^0_{W,Z} & \longrightarrow & M^0_{W,Z} Z_v^{1/2}.
\end{eqnarray}
The term linear in the Higgs field $H$ in the lagrangian
\begin{equation}
  T^0 H^0 = \frac{M^0_W s^0_W}{e_0} (M^0_H)^2 Z_v^{1/2} (Z_v-1) H^0,
\end{equation}
is then used to determine $Z_v$, through the requirement that tadpoles
are canceled. It can be proved \cite{Awramik:2002vu,appelquist} that the bare
masses are gauge invariant in this case.

It should be stressed that this procedure can be advantageous at
higher orders, since the number of diagrams is strongly reduced by
using 1PI Green functions. For the muon decay, there are twice as many
diagrams if tadpoles are introduced. If a treatement per diagram is
required ({\it e.g.} expansion), the calculation time is strongly
correlated to the number of diagrams and substantial differences in
execution can be observed.

Having defined the method of factorization and renormalization, we
turn to the calculation of the respective diagrams. As noticed already
above, all of the bare diagrams are of the vacuum type, and for these
a reduction procedure supplied with analytical formulae
is known \cite{Davydychev:1992mt}. The situation is
slightly more complicated with the on-shell two point functions that
are needed for the on-shell definition of masses. The strategy adopted
in this work consists of two algorithms. The first one
\cite{Weiglein:hd} reduces the tensor integrals to scalar ones and
uses topological symmetries to lead to as small a set of basic
integrals as possible. The integrals obtained in this way are now
subject to numerical evaluation. We used the efficient one dimensional
representation given in \cite{Bauberger:1994by} and extended its
implementation ({\bf S2LSE}) to work with software emulated quadruple precision.

Several tests have been applied to the numerics. First of all, the $Z$
boson propagator has been completely tested by means of the low
momentum expansion, which is valid in this case, as all diagrams are
below threshold. A 6 digit coincidence has been found at tenth
order. The $W$ propagator required the use of large mass expansions,
since several diagrams are at threshold, but the same precision as
previously was also obtainable. Finally both propagators were tested
against the result for the $\overline{\rm MS}$ to on-shell mass
relations of the two gauge bosons given in
\cite{Jegerlehner:2001fb}. This was also considered as a test of the
gauge invariance of the mass counterterms.

Concerning gauge invariance, all of the calculations apart from the
two two loop mass counterterms, have been performed in the general
$R_\xi$ gauge with three parameters and the cancellation of the
dependence on these parameters has been explicitely verified.

\begin{figure}
\psfig{figure=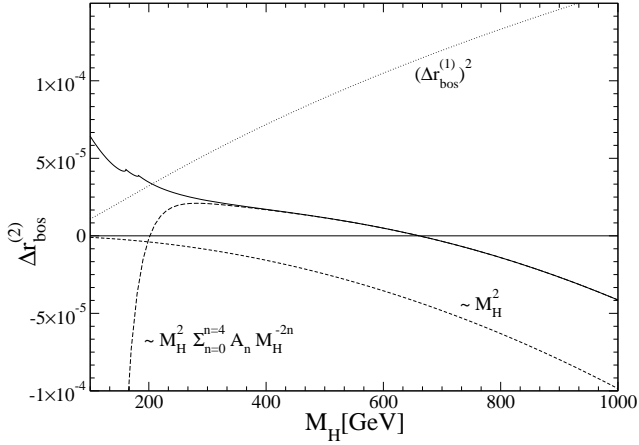,width=6cm,angle=270}
\caption{\label{glowny} The full result for the bosonic corrections to
$\Delta r$ (solid line) against the large mass expansion with leading
term and the square of the one loop result.}
\end{figure}

The full result is given in
Fig.~\ref{glowny} as function of the Higgs boson mass, for the
following values of the input parameters 
\begin{eqnarray}
  \label{parameters}
  && \alpha^{-1} = 137.03599976, \\
  && M_W = 80.423 \mbox{ GeV}, \nonumber \\
  && M_Z = 91.1876 \mbox{ GeV}. \nonumber
\end{eqnarray}
On the same plot, the leading term of the large Higgs boson mass
expansion \cite{Awramik:2002wn,Halzen:1991ik} and the square of the
one loop result are shown to contrast the previous estimates with the
full result.

Our result has also been reexpanded in the large Higgs boson mass
for comparison with \cite{Onishchenko:2002ve} using the
$\overline{MS}$ value and the expansion of the translation formulae
given in \cite{Jegerlehner:2001fb}. The five term expansion is also
shown in Fig.~\ref{glowny}.

It turns out that the variation of the result with the $W$ mass within
the experimental error bars of $\pm 33$ MeV \cite{Parkes:2002vv} is
negligible. The correction can be 
described to $5\%$ accuracy outside of the double threshold region
for Higgs boson masses ranging from 100~GeV to 1~TeV, with the following formula
\begin{equation}
  a+b \Delta+c_1 L+c_2 L^2,
\end{equation}
where
\begin{equation}
\Delta = \left(\frac{M_H}{100\; \mbox{GeV}}\right)^2-1,
\end{equation}
\begin{equation}
L = \log \left[ \left( \frac{M_H}{100\; \mbox{GeV}}\right)^2 \right],
\end{equation}
and the coefficients are
\begin{eqnarray}
a &=& 6.385 \times 10^{-5}, \nonumber \\
b &=& -1.110 \times 10^{-6}, \nonumber \\
c_1 &=& -2.870 \times 10^{-5}, \nonumber \\
c_2 &=& 6.445 \times 10^{-6}.
\end{eqnarray}

\begin{figure}
\psfig{figure=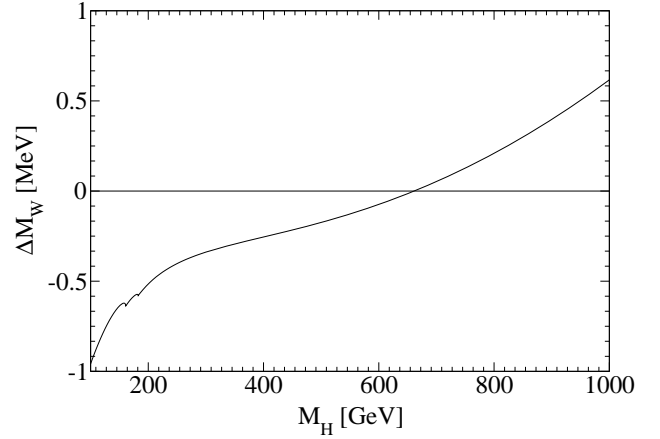,width=6cm,angle=270}
\caption{\label{WMassShift} 
$W$ boson mass shift generated by two loop bosonic corrections.}
\end{figure}

The effect on the prediction of the $W$ boson mass from the muon decay
lifetime is small and can be obtained from an expansion. In fact $6\%$
accuracy is guaranteed by estimating the resulting mass shift from the formula
\begin{equation}
  \label{DMWEQ}
  \Delta M_W = -1.491\times 10^{4} \Delta r^{(2)}_{bos}\mbox{ [MeV]}.
\end{equation}
The respective plot is shown in Fig.~\ref{WMassShift}.

In conclusion, the calculation of the two loop bosonic corrections to
the muon decay lifetime as parametrized by $\Delta r$ has been
reviewed. The crucial steps of the factorization and matching have
been described together. A simple formula describing the correction
for a large range of Higgs boson masses has been given.

M. C. would like to thank the Alexander von Humboldt foundation for
fellowship. This work was supported in part by the European
Community's Human Potential Programme under contract
HPRN-CT-2000-00149 Physics at Colliders, and by the KBN Grants
5P03B09320 and 2P03B05418.

\end{document}